\documentclass{elsart}

\usepackage{graphicx}
\usepackage{amssymb}
\usepackage{amsmath}

\begin{document}

\begin{frontmatter}

% Title, authors and addresses

% use the thanksref command within \title, \author or \address for footnotes;
% use the corauthref command within \author for corresponding author footnotes;
% use the ead command for the email address,
% and the form \ead[url] for the home page:
% \title{Title\thanksref{label1}}
% \thanks[label1]{}
% \author{Name\corauthref{cor1}\thanksref{label2}}
% \ead{email address}
% \ead[url]{home page}
% \thanks[label2]{}
% \corauth[cor1]{}
% \address{Address\thanksref{label3}}
% \thanks[label3]{}

\title{Spatio-temporal conjecture for diffusion}

% use optional labels to link authors explicitly to addresses:
%\author[label1,label2]{}
\author{Mendeli H. Vainstein, }
\author{Rafael Morgado and}
\author{Fernando A. Oliveira}
\ead{fao@fis.unb.br}
\address{Institute of Physics and International Center of Condensed
Matter Physics, University of Bras\'{\i}lia, CP 04513, 70919-970,
Bras\'{\i}lia-DF, Brazil}
% \address[label1]{}
% \address[label2]{}

\author{}

\address{}

\begin{abstract}
% Text of abstract

We present here a conjecture about the equivalence between the noise density of states of a system governed by a generalized Langevin equation and the fluctuation in the energy density of states in a Hamiltonian system. We present evidence of this for a disordered Heisenberg system.

\end{abstract}

\end{frontmatter}

% main text

\section{Introduction}
\label{sec.introduction}

The connection between time and space has been always a matter of discussion in physics and philosophy. 
One can say that a major achievement  last century was the birth of Einstein's theory of relativity which gives equal weight to both.
However, when one looks at  highly disordered systems one seems so far away from such questions that they apparently belong to another world. In this work, we try to present a conjecture which, despite its  empirical character, may be useful to understanding some basic phenomena in complex structures.  

A fundamental question in a physical system is how energy, matter and information propagate.  For systems which present diffusion, more than a century of investigation may lead us to the conclusion that most of the important aspects are known.
However, if we ask simple questions such as ``How do spin waves diffuse in a Heisenberg system with correlated disorder?'', ``How do electrons behave in an irregular lattice?'', or ``How does a ratchet device work?'', it takes a short time to realize that these unanswered problems are related to diffusion. In this way, we can say that diffusion is still a modern and present problem in all condensed matter physics.
 
In order to try to answer  questions as those above, we shall put them in two different frameworks:  First, let us consider a system  
 described by a generalized Langevin equation~\cite{Kubo91} 
of the form

\begin{equation}
\frac{dA(t)}{dt}=-\int _{0}^{t}\Gamma (t-t')A(t')dt'+F(t),
\label{GLE}
\end{equation}
 where $F(t)$ is a stochastic noise subject to the conditions $\langle F(t)A(0)\rangle =0$, $\langle F(t)\rangle =0$, and the Kubo fluctuation-dissipation theorem, namely $C_{F}(t)=\langle F(t)F(0) \rangle=\langle A^{2}\rangle_{eq}\Gamma (t)$ (FDT).  Here, the brackets $\langle\rangle$ mean ensemble averages.
In order to study diffusion, let us now define the variable $x(t)$ as
\begin{equation}
x(t)=\int _{0}^{t}A(t')dt',
\label{3}
\end{equation}
 and suppose its mean square value has the  asymptotic behavior
\begin{equation}
\lim _{t\rightarrow \infty }<x^{2}(t)>\sim t^{\alpha }.
\label{x2}
\end{equation}
 For normal diffusion $\alpha =1$; we have subdiffusion for $\alpha <1$
and superdiffusion for $\alpha >1$.
 Notice that if $A(t)$ is the momentum of a particle, $x(t)/m$ is its position.
 Recently~\cite{Morgado02}, it was proved that if the Laplace transform of the memory behaves as                                           
\begin{equation}
  \tilde{\Gamma }(z\rightarrow 0) \sim z^{\nu},
\label{Gaz}
\end{equation}
then
\begin{equation}
\alpha= \nu+1.
\label{alpha}
\end{equation}
Therefore, for this kind of problem, a great deal of achievement has been obtained, and the kind of diffusion can be immediately predicted without tedious calculations. 
This system presents noise with an associated memory function given by
 \begin{equation}
\label{Gamma}
\Gamma (t)=\int \rho_n (\omega )\cos (\omega t)d\omega.
\end{equation}
Here $\rho_n(\omega)$, is the noise density of states (NDS). This result can be obtained from the FDT~\cite{Morgado02,Costa03}.
We shall say that such a system presents \emph{temporal disorder.}

A second class of systems is composed by those which present \emph{spatial disorder}. They have been thoroughly investigated in the last half century~\cite{Anderson58}, nevertheless, some questions concerning localization or diffusion still remain open. Let us consider as an example the Heisenberg chain~\cite{Evangelou92,Moura02} 
 \begin{equation}
\label{Heisenberg}
H=-\sum _{l=1}^{N}J_{l}\mathbf{S}_{l}\cdot \mathbf{S}_{l+1},
\end{equation}
 where $S=1/2$. Here, $J_l$ is the exchange integral at the site $l$. Equivalently, we could consider the disordered harmonic chain~\cite{Moura03}  or even the Anderson model~\cite{Moura98}.

   Can we predict the properties of those system in the same way we do for the GLE? The answer is partially yes, partially no. The conjecture we present here, being valid, will help to answer those questions.

This work is organized as follows: 
 In Sec.~\ref{sec.conjecture}, we present the spatio-temporal conjecture.
In Sec.~\ref{sec.consideration}, we outline some considerations about the conjecture. In Sec.~\ref{sec.Heisenberg} we show that the conjecture works for the quantum Heisenberg chain. 
Finally we  conclude the work in Sec.~\ref{sec.conclusion}.
                                                                               
\section{The Conjecture}
\label{sec.conjecture}

   Consider a system which presents fluctuations in its energy density of states $D(E)$; let us call $\rho_F(E)$ the fluctuation density. Then
\begin{equation}
\rho_F(E) = \rho_n(E).
\label{conjecture}
\end{equation}
If this is true, then  $\rho_F$ can be introduced in Eq.~(\ref{Gamma}) to obtain the diffusive exponent $\alpha$. This is our conjecture.
 Notice that by energy density of states we usually refer to one particle density of states, or, see Sec.~\ref{sec.Heisenberg}, to the density of states for the spin wave dynamics. For any situation one shall define the appropriated $D(E)$.  This conjecture was somehow advanced by Morgado \emph{et al.}~\cite{Morgado02}. However, it is stated here in a clear and precise manner. 

\section{Considerations about the conjecture}
\label{sec.consideration}

   A fundamental aspect of the stochastic process is the noise. Consequently, no one will question the result presented in Eq.~(\ref{alpha}); in any case it can be easily verified in reference~\cite{Morgado02}.
The question is why is the fluctuation important and not the density itself, since we have learned from statistical mechanics that $D(E)$ will give us the main information we need? 
We may advance an insight: \emph{the relaxation process may depend on all the fluctuations  the system presents, i.e., the fluctuation in energy levels and in the dynamical variables}.
 For most systems the fluctuation in the density of states will follow the law of the great numbers $N^{-1/2}$ and they are irrelevant. Nevertheless, for some mesoscopic systems with  highly correlated disorder, or due to some other reason with important fluctuations, these will be dominant for the relaxation process.

The relations between time and space are common even in situations where we do not expect them. For example, in systems which have broken space symmetry due to the presence of surfaces or disorder, time reversal symmetry allows the persistence of some space reversion symmetries~\cite{Oliveira81}.

\section{Ferromagnetic Heisenberg disordered chain}
\label{sec.Heisenberg}

In this section we discuss a concrete case where the conjecture works. We present here an illustrative situation; details will be given elsewhere~\cite{Vainstein03}. Let us consider the chain described by Eq.~(\ref{Heisenberg}) with couplings given by 

\begin{equation}
J_l = J_0+\sum _{k=1}^{N/2}(Ck^{-\beta})^{1/2}\cos (\pi kl/N+\phi _{k}),
\label{A0}
\end{equation}
 where $C$ is a constant, the exponent $\beta$ controls the strength of the correlation, and the angles $\phi$ are random phases in the interval $0 \le \phi < 2\pi$. Here $J_0$ shifts all $J_l$, so that $J_l > 0$, in order to keep the chain ferromagnetic. By direct numerical integration~\cite{Moura02}, it has been shown that for $0 < \beta < 1$ the spin waves are delocalized and present a superdiffusive behavior with $\alpha=1.5$; for $1 < \beta < 2$, the system presents ballistic motion, i.e., $\alpha=2$. 
 
\begin{figure}
\centering
\includegraphics[height=7cm,width=5cm,angle=270]{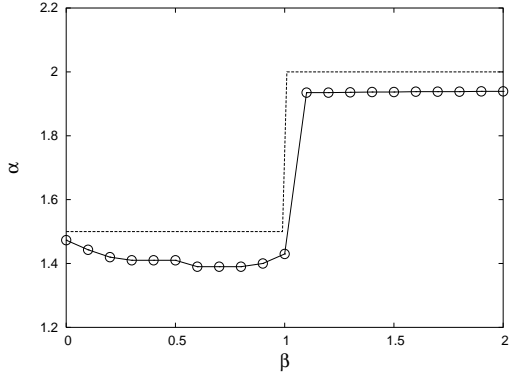}
\caption{Numerical data for the diffusion exponent $\alpha $ as function of $\beta $ averaged over $900$ chains.}
\label{fig.1} 
\end{figure}

 We use a renormalization process~\cite{Moura02} to obtain the density of states $D(E)$. We then define the fluctuation in the density of states as
\begin{equation}
\rho_F^2(E) = \langle (D(E)-\langle D(E) \rangle)^2 \rangle.
\end{equation} 
Here, the average is done over many chain realizations.  We use the conjecture Eq.~(\ref{conjecture}), and the relations (\ref{Gaz}-\ref{Gamma}) to obtain $\alpha$. 
  In Fig.~\ref{fig.1}, we show $\alpha$ as a function of $\beta$.  The line is the result from numerical integration while the points are from the conjecture. The agreement is remarkable.
Consequently, for the Heisenberg chain, the conjecture works! For the harmonic chains  and the Anderson model with similar disorder, the preliminary results also reproduce the result obtained by numerical integration.

%\section{Limits of the conjecture\label{sec.limit}}

   A complete description of a system such as the Heisenberg chain above is not fully possible. For example, one cannot discover the value of $\alpha$ for a given value of $\beta$, unless we compute $D(E)$ and then $\rho_F(E)$.
 At this point, we may argue that if we do not know $\rho_n(\omega)$, then we also do not know how to compute $\alpha$; therefore, we place the two approaches at the same level.
The \emph{temporal approach}, however, sounds simple, since there is apparently no restriction to the values of $\alpha$ one can obtain. 
The limits of the diffusion we have are easily understood, since spin waves have well defined dispersion relations,
 just like sound waves, solitons, or any other excitation. They all have well defined density of states. The question now is why two regimes? 
Again, the determining rule of the fluctuations for the diffusion helps to explain this. 
The same dispersion relations present very different fluctuations as the correlation in the disorder changes. 
Notice that the intensity of the fluctuations is not important, as long as it is relevant: in the Eq.~(\ref{Gaz}), we see that only the exponent $\nu$ is important for the determination of $\alpha$. This fact makes the conjecture acceptable even when the FDT does not work~\cite{Costa03}. 
As we change $\beta$ in the interval $0 < \beta < 1$, the system preserves  the same law with the fluctuations changing their intensity. 
For $\beta \sim 1$, it can no longer support the same dispersion relation and fluctuations, so it changes to a new ``order'' with new fluctuations. 
One characteristic of ballistic motion, $1 < \beta < 2$, is the absence of fluctuations in the low density region~\cite{Costa03}. 
Throughout the region  $0 < \beta < 2$, the density of states has the same law $D(E) \sim E^{-1/2}$, within the calculation's precision~\cite{Moura02}.
 That is how the system preserves its dispersion relation for the spin waves, while  it responds to different disorder. This shows how important the fluctuations are in the determination of the diffusive process. 

%Finally, it is important to be within the limit of the conjecture \emph{ an equivalence between temporal fluctuations (noise) and spatial fluctuations (fluctuations in the energy density of states)}.
% That is precisely what Eq.~(\ref{conjecture}) establishes. For systems where fluctuations do not play an important role it will not be a surprise if the conjecture does not hold.  

\section{Conclusion}
\label{sec.conclusion}

We have discussed the diffusive properties of a dynamical variable $A(t)$ and we proposed the conjecture  that there is an equivalence between the noise in a GLE and the fluctuation in the energy density of states, in such way that stochastic and Hamiltonian methods can be interchanged to describe diffusion. The conjecture may appear either naively simple or wrong.
 However, it deserves full attention due to two reasons. From the result of Sec.~\ref{sec.Heisenberg}, we see that it works well for a very important and complex system: the quantum disordered Heisenberg chain.  
Second, if one can classify systems where it works and where it does not, it would create a great step forward, and
 it would promote new research in the direction of universal diffusive behavior.

ACKNOWLEDGEMENTS:

We thank F.~A.~B.~F. de Moura and M.~D. Coutinho-Filho for very important discussions. This work was supported by CAPES, CNPq, FINEP and FINATEC.

% The Appendices part is started with the command \appendix;
% appendix sections are then done as normal sections
% \appendix

% \section{}
% \label{}

%\begin{thebibliography}{00}
% BibTeX users please use
\bibliographystyle{elsart-num}
%\bibliography{conjecture}
% \bibitem{label}
% Text of bibliographic item

% notes:
% \bibitem{label} \note

% subbibitems:
% \begin{subbibitems}{label}
% \bibitem{label1}
% \bibitem{label2}
% If there is a note, it should come last:
% \bibitem{label3} \note
% \end{subbibitems}

%\bibitem{}

%\end{thebibliography}

\end{document}